# ВИЗНАЧЕННЯ ФУНКЦІОНАЛЬНОГО СТАНУ ФРУКТОВИХ ПРОДУКТІВ ЗА ПАРАМЕТРАМИ ЇХ ЕЛЕКТРИЧНОГО ІМПЕДАНСУ

***Гусева О. В., к.т.н.,доц.; Мосійчук В. С., к.т.н., доц.; Тимошенко Г. В., аспірант; Шарпан О .Б. д.т.н.,проф.***

*Національний технічний університет України «Київський політехнічний інститут», Київ, Україна*

# DEFINITION OF FUNCTIONAL CONDITION OF THE PRODUCT ACCORDING TO THE PARAMETERS OF THEIR COMPLEX IMPEDANCE

***Guseva O.V., Mosiychuk V. S., Timoshenko G. V., Sharpan O. B.***

*National Technical University of Ukraine, Kyiv Politechnic Institute, Kiev, Ukraine*

## Вступ

Для визначення функціонального стану (ФС) продуктів та оцінки факторів, що впливають на їх якість, часто використовують значення їх електропровідності. Зокрема, значення електропровідності (або електричного опору) визначають, оцінюючі структуру та стан клітин і міжклітинного простору. З функціональним станом продуктів безпосередньо пов’язана гідратація та зміна структури, наприклад, внаслідок значних деструктивних процесів, зміни у будові клітин та у хімічному складі кліткових рідин і мембран. На параметри комплексного опору біопродуктів (овочів, фруктів) також впливає їх структура (кора та м’якоть плоду, напрямок волокон), форма, а також гідратація та температура [1].

Відомі застосування подібних способів оцінки ФС для визначення зрілості плодів шляхом занурення електродів в плід та вимірювання його електропровідності в діапазоні частот 400-500 Гц [2], або для визначення часу виходу плодів картоплі з стану спокою для визначення умов початку її проростання [3]. Проте в даних дослідженнях не враховувалася реактивна складова електричного імпедансу під час вимірювань, оскільки її сприймали як таку, що сприяє підвищенню похибки [2]. Слід зазначити, що здебільшого оцінювався лише модуль імпедансу і додатково висувалися вимоги до товщини зразків. Наприклад в [4] встановлена мінімальна фіксована відстань між електродами 60 мм, оскільки на тонкіших зразках опір визначається нульовим. Також, для оцінювання неоднорідності структур біооб’єктів необхідно визначати параметри їх імпедансу у розширеній смузі частот [6].

**Метою статті** є визначення критеріїв оцінки функціонального стану продуктів за всіма параметрами частотних характеристик їх імпедансу.

## Методика досліджень

Для дослідження функціонального стану фруктів за всіма параметрами електричного імпедансу були підготовлені групи плодів яблук, груш та лимонів. З плодів виготовлялися зразки, котрі надалі використовувалися для вимірювання модуля та фази їх імпедансу. Підготовлені таким чином зразки плодів піддавались деструктивним змінам в лабораторних умовах за кімнатної температури протягом одного тижня. Досліджувалась залежність зміни параметрів імпедансу від дегідратації та зміни структури клітин (гниття).

Для оцінки можливостей і перспектив розвитку діагностичних методик визначення функціонального стану використовувались одночастотний [8] та тричастотний ТОР-М1 [7] вимірювачі параметрів біоімпедансу. Особливістю апарата ТОР-М1 є те, що він забезпечує визначення усіх складових (модуля *Z*, фази φ, активної *R* та реактивної *X*) імпедансу на частотах 20 кГц, 100 кГц та 500 кГц [7, 8]. За допомогою вимірювача [6] параметри імпедансу отримувались додатково на частоті 250 кГц. Вимірювання повного імпедансу фруктів здійснювалось за чотирьохелектродною методикою з використанням електродів, в яких відстань між струмовим і потенціометричним електродом в кожній парі конструктивно фіксована і дорівнює 10 мм. Площа кожного струмового електрода становить 1,76 см$^2$, а потенційного, що конструктивно розташовується навколо струмового – 23,35 см$^2$. Матеріал електродів — нержавіюча сталь. З метою усунення впливу розмірів та форми досліджуваних об'єктів для вимірювання підготовлювалися препарати різних фруктів (груша, яблука, лимон) у вигляді поперечного та повздовжнього зрізів товщиною 35 мм. Контактна рідина не застосовувалась у зв'язку з особливостями досліджуваного об'єкту. Проводилися виміри імпедансу свіжого продукту, через три та через шість діб. Протягом цього часу фрукти зазнавати змін функціонального стану (псування продукту), що проявлялося процесами гниття та дегідратації. У цьому разі умови протягом перших трьох днів сприяли процесам гниття, а протягом інших трьох — дегідратації.

## Результати досліджень

Вплив змін функціонального стану різних продуктів під час їх псування на параметри повного імпедансу на частотах 20, 100 і 500 кГц приведені у табл. 1 – 7. Значення, що приведені у таблицях, отримані шляхом усереднення 256 послідовних вимірювань, що забезпечує зменшення впливу флуктуацій. Похибка вимірювань не перевищувала 5 %.

Табл. 1 – Процес гниття яблука

| стан | Модуль (Ом) | | | Фаза (град) | | |
|---|---|---|---|---|---|---|
| | *20kHz* | *100kHz* | *500kHz* | *20kHz* | *100kHz* | *500kHz* |
| 25% | 716 | 542 | 375 | -18,3 | -16,6 | -8,4 |
| 3 дні, 45% | 645 | 306 | 229 | -24,9 | -20,7 | -6,9 |
| 6 днів, 90% | 287 | 462 | 411 | -63,7 | -75,5 | -46,0 |

В табл. 1 приведена динаміка зміни значення модуля і фази імпедансу поперечного зрізу яблука товщиною 35 мм, в якому спостерігалася ділянка площі, враженої процесами гниття. При цьому на початковому етапі площа цієї ділянки складала 25% і збільшилась через 3 дні до 45%, а через 6 днів — до 90%. Значення модуля і фази імпедансу зрізу яблука товщиною 35 мм залежно від зміни його функціонального стану показані в табл. 2 для повздовжнього та в табл. 3 для поперечного зрізу.

Табл. 2 – Поздовжній зріз яблука

| стан | Модуль (Ом) | | | Фаза (град) | | |
|---|---|---|---|---|---|---|
| | *20kHz* | *100kHz* | *500kHz* | *20kHz* | *100kHz* | *500kHz* |
| свіжий | 670 | 344 | 245 | -43.3 | -25.3 | -10.8 |
| 3 дні | 974 | 659 | 388 | -43.1 | -26.4 | -16.5 |
| 6 днів | 6950 | 2240 | 758 | -31.6 | -38.4 | -55.2 |

Табл. 3 – Поперечний зріз яблука

| стан | Модуль (Ом) | | | Фаза (град) | | |
|---|---|---|---|---|---|---|
| | *20kHz* | *100kHz* | *500kHz* | *20kHz* | *100kHz* | *500kHz* |
| свіжий | 739 | 260 | 112 | -48 | -31.1 | -7.1 |
| 3 дні | 974 | 595 | 334 | -47.2 | -30.8 | -12.6 |
| 6 днів | 5950 | 1955 | 731 | -24.1 | -20.5 | -34.6 |

В табл. 4 і 5 наведені значення модуля і фази імпедансу повздовжнього (табл. 4) та поперечного (табл. 5) зрізу груші товщиною 35 мм залежно від зміни його функціонального стану.

Табл. 4 – Поздовжній зріз груші

| стан | Модуль (Ом) | | | Фаза (град) | | |
|---|---|---|---|---|---|---|
| | *20kHz* | *100kHz* | *500kHz* | *20kHz* | *100kHz* | *500kHz* |
| свіжий | 148 | 0 | 0 | 0 | -20,4 | -15,2 |
| 3 дні | 403 | 50 | 0 | -38,3 | -30,4 | -10,5 |
| 6 днів | 346 | 0 | 0 | -27,6 | -25 | -4,5 |

Табл. 5 – Поперечний зріз груші

| стан | Модуль (Ом) | | | Фаза (град) | | |
|---|---|---|---|---|---|---|
| | *20kHz* | *100kHz* | *500kHz* | *20kHz* | *100kHz* | *500kHz* |
| свіжий | 885 | 323 | 144 | -47.5 | -33,8 | -10,8 |
| 3 дні | 226 | 69 | 14 | -43,6 | -30,2 | -5,3 |
| 6 днів | 1312 | 1072 | 516 | -11,0 | -35,9 | -41,0 |

В табл. 6 і 7 наведені значення модуля і фази імпедансу повздовжнього (табл. 6) та поперечного (табл. 7) зрізу лимона товщиною 35 мм залежно від зміни його функціонального стану.

Табл. 6 – Поздовжній зріз лимону

| стан | Модуль (Ом) | | | Фаза (град) | | |
|---|---|---|---|---|---|---|
| | *20kHz* | *100kHz* | *500kHz* | *20kHz* | *100kHz* | *500kHz* |
| свіжий | 503 | 3 | 0 | -45,1 | -32,6 | -7,4 |
| 3 дні | 553 | 68 | 0 | -47,9 | -43,1 | -6,9 |
| 6 днів | 235 | 99 | 0 | -9,0 | -21,7 | -19,8 |

Табл. 7 – Поперечний зріз лимону

| стан | Модуль (Ом) | | | Фаза (град) | | |
|---|---|---|---|---|---|---|
| | *20kHz* | *100kHz* | *500kHz* | *20kHz* | *100kHz* | *500kHz* |
| свіжий | 974 | 263 | 158 | -37,4 | -39,1 | -24,5 |
| 3 дні | 308 | 97 | 74 | -35,5 | -28,4 | -7,4 |
| 6 днів | 68 | 63 | 0 | -22,9 | -15,6 | -55,0 |

## Обробка та аналіз результатів

Для дослідження ФС продуктів бажано на основі експериментальних даних відтворити їх амплітудно-частотні і фазочастотні характеристики в досліджуваному частотному діапазоні [5]. Для цього об'єкт дослідження доцільно представити у вигляді двополюсника на основі простої схеми, що складається з зосереджених елементів [9].

У разі моделювання двополюсника необхідно спочатку відтворити частотну узагальнену функцію комплексного опору:

$$Z(p) = \frac{A_0 + A_1 p + \ldots + A_N p^N}{1 + B_1 p + \ldots + B_M p^M}, \quad N \le M, \tag{1}$$

тобто відтворити коефіцієнти $A_n$, $n = 0 \ldots N$, $B_m$, $m = 0 \ldots M$, що повинні задовольняти умові додатньої визначеності. Далі визначають параметри двополюсника за відомою топологією (наприклад, однієї з схем на рис. 1).

Вихідними даними для визначення функції (1) можуть бути значення активної та реактивної складових імпедансу або значення повного опору на декількох частотах. В першому разі вираз (1) для схем на рис.1,а,б можна записати у вигляді

$$\frac{A_0 + jA_1\omega - A_2\omega^2}{1 + jB_1\omega - B_2\omega^2 - jB_3\omega^3} = R(\omega) + jX(\omega). \qquad (2)$$

Приводячи (2) до загального знаменника та враховуючі властивості комплексних чисел, отримаємо на кожній частоті два рівняння:

$$\begin{cases} A_0 - \omega^2 A_2 + \omega X(\omega) B_1 + \omega^2 R(\omega) B_2 - \omega^3 X(\omega) B_3 = R(\omega), \\ \omega A_1 - \omega R(\omega) B_1 + \omega^2 X(\omega) B_2 + \omega^3 R(\omega) B_3 = X(\omega). \end{cases} \qquad (3)$$

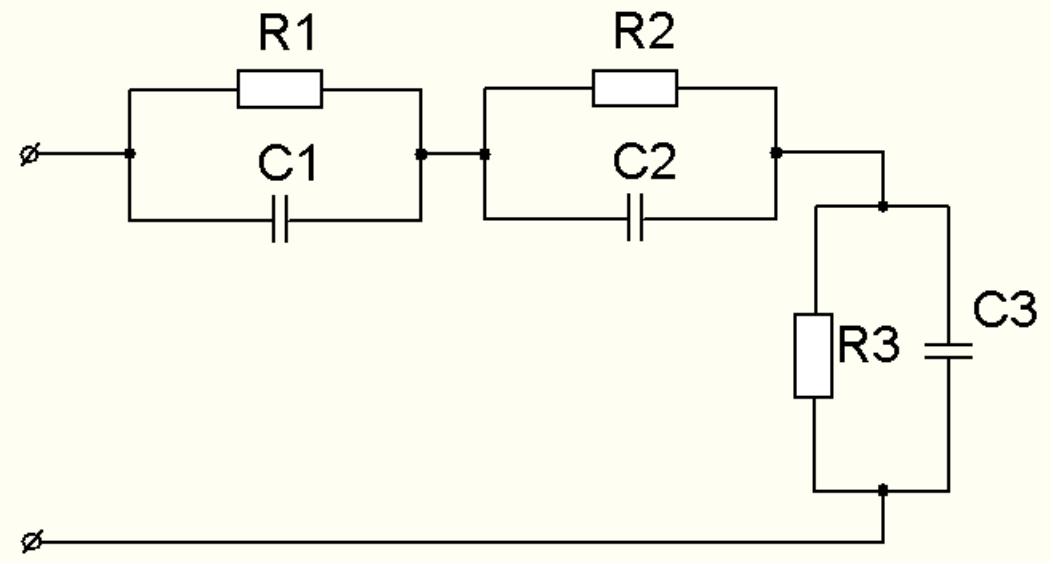

$$\dot{Z}(\omega) = \frac{R_1}{1 + j\omega R_1 C_1} + \frac{R_2}{1 + j\omega R_2 C_2} + \frac{R_3}{1 + j\omega R_3 C_3};$$

***а***

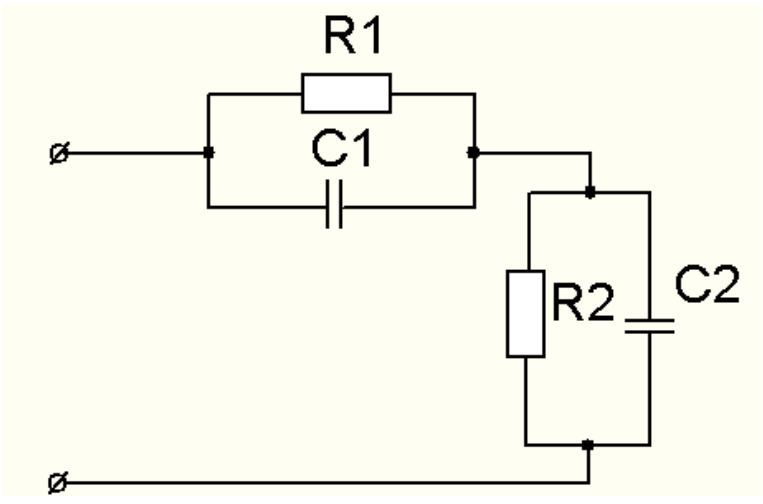

$$\dot{Z}(\omega) = \frac{R_1}{1 + j\omega R_1 C_1} + \frac{R_2}{1 + j\omega R_2 C_2};$$

***б***

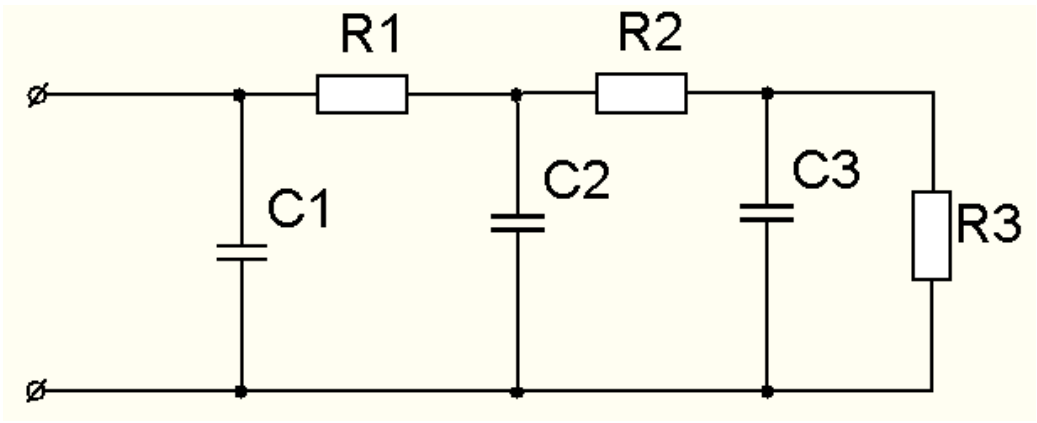

$$\dot{Z}(\omega) = \cfrac{1}{j\omega C_1 + \cfrac{1}{R_1 + \cfrac{1}{j\omega C_2 + \cfrac{1}{R_2 + \cfrac{1}{j\omega C_3 + \cfrac{1}{R_3}}}}}};$$

***в***

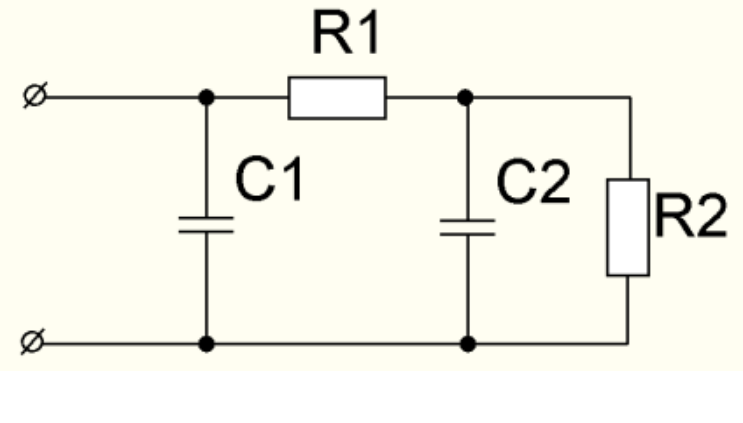

$$\dot{Z}(\omega) = \cfrac{1}{j\omega C_1 + \cfrac{1}{R_1 + \cfrac{1}{j\omega C_2 + \cfrac{1}{R_2}}}};$$

***г***

Рис. 1 — Моделювання двополюсників за частотною характеристикою імпедансів фруктів

Далі в табл. 8 приведені результати моделювання для випадків представлення об'єкту дослідження різними топологіями (моделями) (рис.1).

Таблиця 8 – Отримані параметри двополюсників для різних фруктів

| Об'єкт | зразок | схема рис.1 | $C_1$, нФ | $R_1$, кОм | $C_2$, нФ | $R_2$, кОм | $C_3$, нФ | $R_3$, кОм |
|---|---|---|---|---|---|---|---|---|
| Яблуко товщиною 35 мм, повздовжній зріз (рис.2) | свіжий | (а) | 0.00214 | 0.237 | 11.169 | 0.102 | 16.029 | 1.464 |
| | | (в) | 0.00214 | 0.237 | 6.584 | 0.283 | 11.145 | 1.282 |
| | 3 дні | (а) | 0.01495 | 0.375 | 3.306 | 0.304 | 12.638 | 10.659 |
| | | (в) | 0.01498 | 0.379 | 2.629 | 0.479 | 10.186 | 10.479 |
| | 6 днів | (а) | 0.04076 | 0.285 | 0.526 | 1.980 | 3.436 | 4.113 |
| | | (в) | 0.03742 | 0.338 | 0.427 | 2.557 | 3.549 | 3.484 |
| Яблуко товщиною 35 мм, поперечний зріз (рис.3) | свіжий | (а) | 0.00098 | 0.210 | 170.15 | 0.010 | 11.412 | 1.317 |
| | | (в) | 0.00098 | 0.210 | 12.231 | 2.644 | 0.618 | 1.338 |
| | 3 дні | (а) | 0.0101 | 0.373 | 6.707 | 0.243 | 8.999 | 3.528 |
| | | (в) | 0.0101 | 0.375 | 3.844 | 0.711 | 6.166 | 3.057 |
| | 6 днів | (а) | 0.00065 | 0.603 | 0.403 | 1.365 | 5.231 | 2.419 |
| | | (в) | 0.00065 | 0.605 | 0.374 | 1.576 | 5.344 | 2.206 |
| Яблуко – площа, вражена процесами гниття (рис.4) | 25% | (а) | 0.00234 | 0.374 | 6.634 | 0.195 | 29.751 | 0.439 |
| | | (в) | 0.00234 | 0.375 | 5.428 | 0.286 | 31.323 | 0.348 |
| | 45% | (а) | 0.00974 | 0.266 | 25.373 | 0.0156 | 13.662 | 0.531 |
| | | (в) | 0.00973 | 0.266 | 8.916 | 0.116 | 6.456 | 0.430 |
| | 90% | (а) | 1.591 | 0.247 | 2.150 | 0.255 | 23.991 | 0.647 |
| | | (в) | 4.880 | 0.0558 | 54.495 | 0.0050 | 76.729 | 0.588 |
| Лимон товщиною 35 мм, поперечний зріз (рис.5) | свіжий | (б) | 0.5478 | 0.1523 | 11.339 | 0.7676 | | |
| | | (г) | 0.5226 | 0.1676 | 11.038 | 0.7526 | | |
| | 3 дні | (б) | 0.0279 | 0.0780 | 32.980 | 0.2849 | | |
| | | (г) | 0.0279 | 0.0782 | 32.968 | 0.2848 | | |
| | 6 днів | (б) | 6.931 | 0.0636 | 341.02 | 4.140 | | |
| | | (г) | 6.793 | 0.0662 | 334.44 | 4.137 | | |
| Груша товщиною 35 мм, поперечний зріз (рис.6) | свіжий | (б) | 0.0655 | 0.2395 | 9.373 | 1.5446 | | |
| | | (г) | 0.0651 | 0.2429 | 9.3283 | 1.5413 | | |
| | 3 дні | (б) | 0.0536 | 0.0477 | 78.563 | 0.2139 | | |
| | | (г) | 0.0536 | 0.0478 | 78.533 | 0.2138 | | |
| | 6 днів | (б) | 0.3166 | 0.4253 | 1.6725 | 1.2351 | | |
| | | (г) | 0.2662 | 0.5943 | 1.6492 | 1.0662 | | |

Частотні характеристики, що побудовані на основі даних табл. 8, наведені на рис.2 – 6.

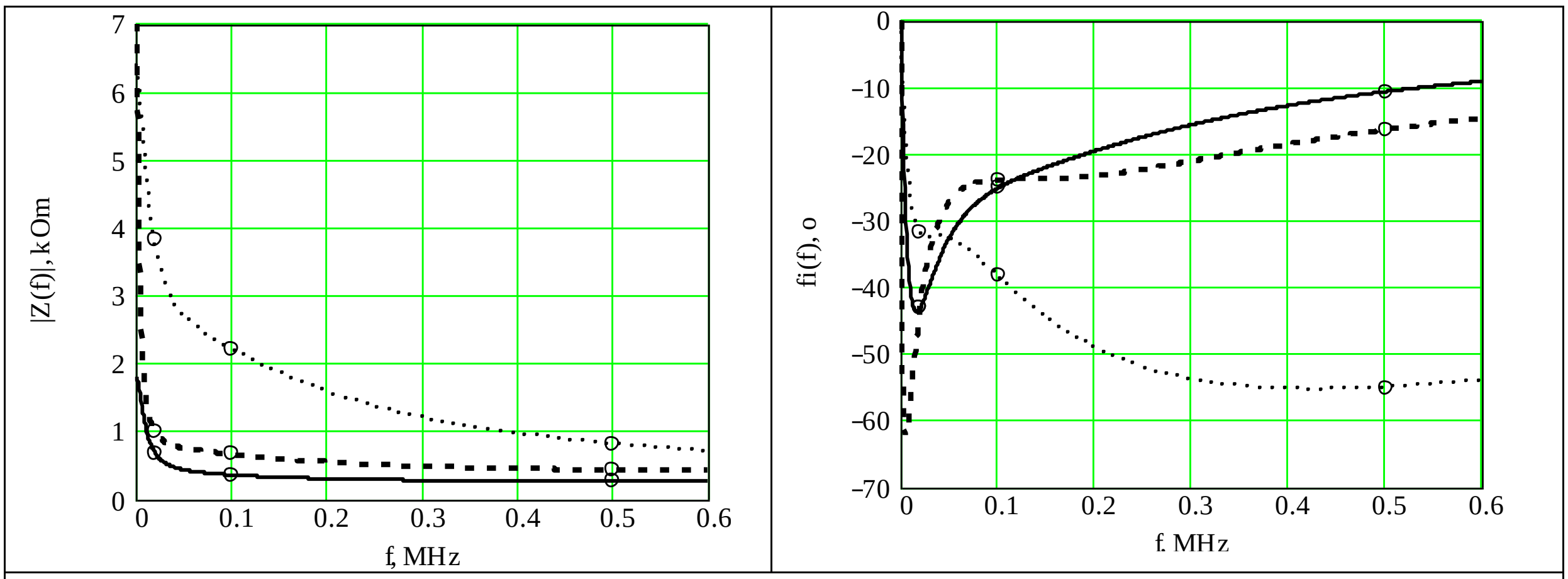

Рис. 2 — Поздовжній зріз яблука товщиною 35 мм

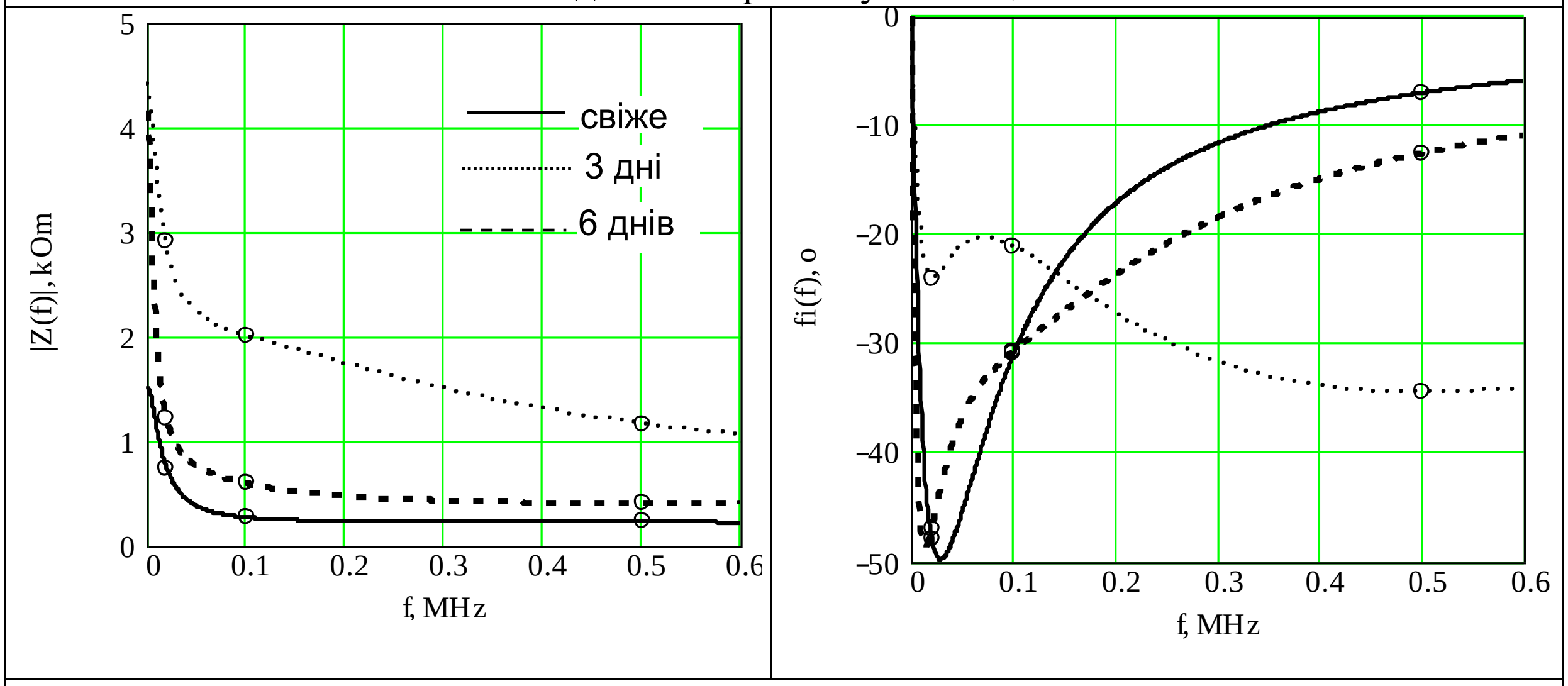

Рис. 3 — Поперечний зріз яблука товщиною 35 мм

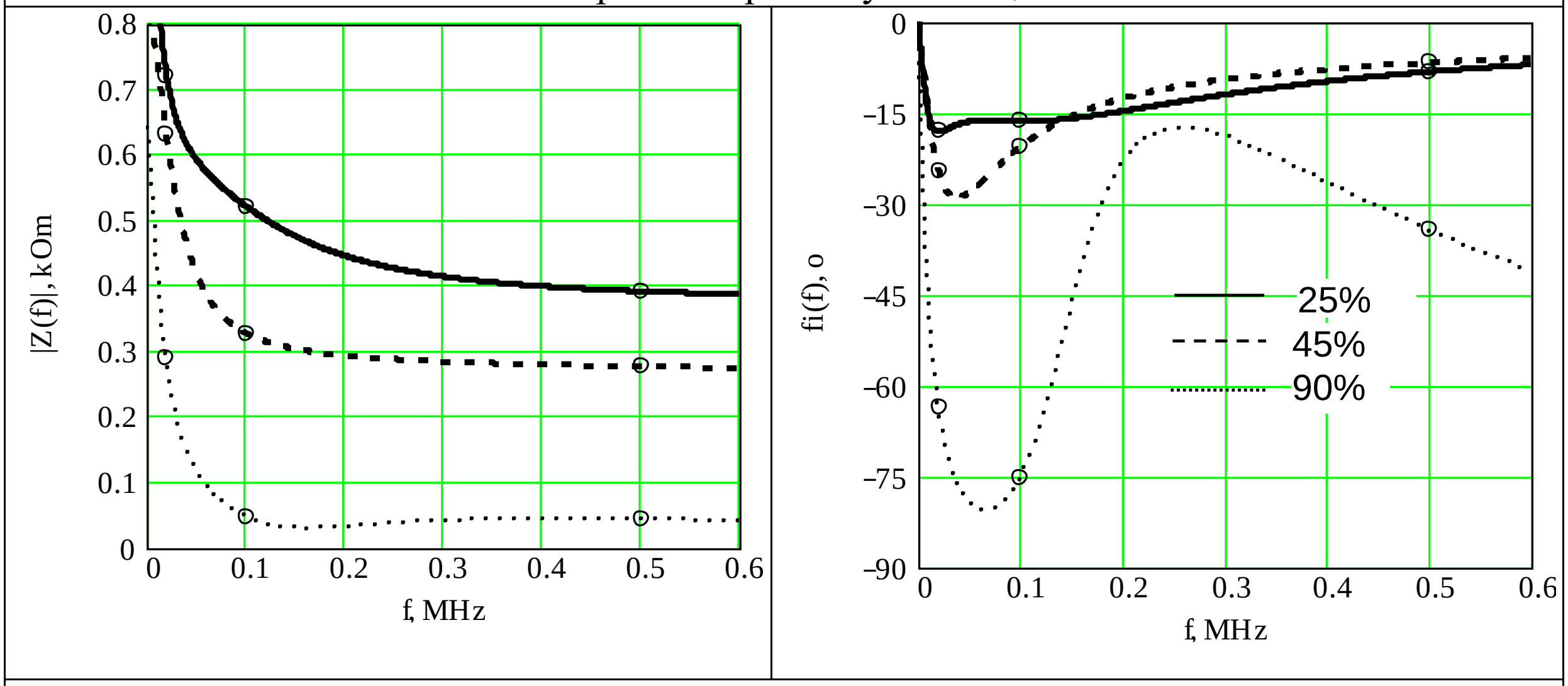

Рис. 4 — Яблуко – площа, вражена процесами гниття

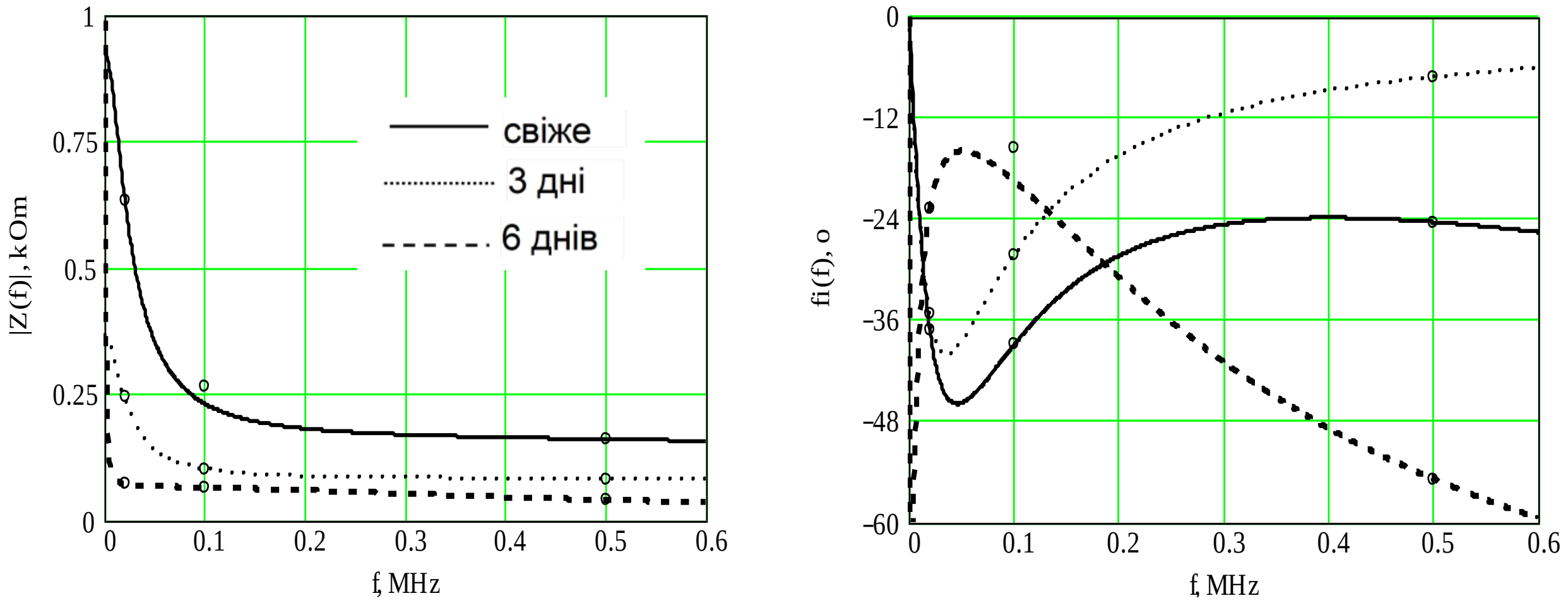

Рис. 5 — Поперечний зріз лимона товщиною 35 мм

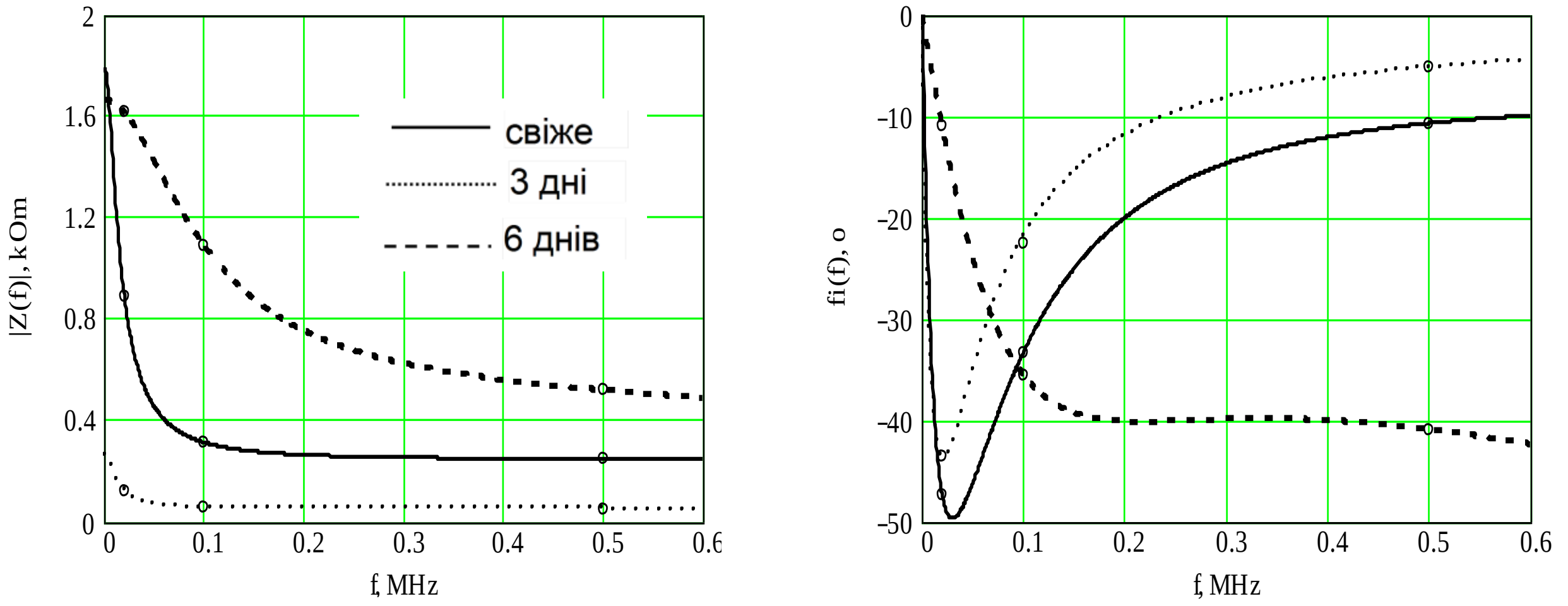

Рис. 6 — Поперечний зріз груші товщиною 35 мм

У таблиці 8 наведені дані для різних моделей (рис.1), що описують АЧХ і ФЧХ (рис. 2 – 6) продуктів під час змін їх функціонального стану та структури. Слід зазначити, що у разі значних змін функціонального стану та структури, доцільно розглядати і зміни у топології моделі.

## Обговорення результатів та висновки

Свіжі та несвіжі фрукти, такі як яблука, можливо розрізняти за значеннями реактивної складової комплексного опору на низьких та високих частотах, наприклад, 20 та 500 кГц. У свіжих фруктах значення різниці фаз між струмом та напругою на низькій частоті більша ніж у несвіжих продуктів майже у два рази, і навпаки, на високій частоті значення різниці фаз свіжих фруктів менше більш ніж у два рази, чим у несвіжих.

Залежно від процесів, що проходять в плодах під час їх псування (дегідратація чи гниття), характерно змінюються і параметри електричного імпедансу на різних частотах. У разі дегідратації значення модуля імпедансу яблука на всіх частотах збільшуються, а під час гниття зменшуються. У лимона модуль імпедансу, навпаки, в процесі дегідратації зменшується, а у груші він в процесі старіння має екстремум.

Фазова характеристика імпедансу в усіх дослідах в процесі старіння кардинально змінює свій характер.

Процеси старіння та псування продуктів проявляються характерною зміною реактивної складової на різних частотах. Рекомендується для діагностики свіжості продуктів застосовувати фазові портрети. Фазові портрети для окремих фруктів та овочів слід формувати емпірично

**Перелік посилань**

*Гусева О. В., Мосійчук В. С., Тимошенко Г. В., Шарпан О. Б.* ***Визначення функціонального стану фруктових продуктів за параметрами їх електричного імпедансу.*** *Розглянуто можливість визначення функціонального стану фруктів за модулем і фазою їх імпедансу, отриманих за вимірюванням на трьох частотах 20, 100 та 500 кГц. Функціональний стан визначався за динамікою параметрів електричного імпедансу під час деструктивних процесів, викликаних дегідратацією та гниття. На основі експериментально отриманих даних на трьох частотах змодельовані зміни амплітудно- та фазочастотної характеристик. Визначено характерні критерії, за якими можливо виконувати оцінку функціонального стану продуктів.*

*Гусева Е. В., Мосийчук В. С., Тимошенко Г. В., Шарпан О. Б.* ***Определение функционального состояния фруктовых продуктов по параметрам их электрического импеданса.*** *Рассмотрена возможность определения функционального состояния фруктов по модулю и фазе их импеданса, измеренных на трех частотах 20, 100 и 500 кГц. Функциональное состояние определялось по динамике изменений параметров электрического импеданса вследствие деструктивных процессов, вызванных дегидратацией и гниения. На основе экспериментально полученных данных на трёх частотах смоделированы изменения амплитудно-частотной и фазочастотной характеристик. Определены характерные критерии, по которым можно проводить оценку функционального состояния свежих и несвежих продуктов.*

*Guseva E. V., Mosiychuk V. S., Timoshenko G. V., Sharpan O. B.*

***Determination of the functional state of the fruits by parameters of the electric impedance.***

*Introduction*. To assess the freshness of various products are often used measuring impedance module. But due to the structure of plant foods diagnostic value should have exactly a complex component of impedance. Article tasked with developing criteria for assessing the functional state of the products subject to a comprehensive component of the impedance. *Research methodology*. To determine the functional status of the fruit were measured module and phase of impedance at the three frequencies of 20, 100 and 500 kHz. Criteria for recognition of functional status determined by the dynamics of changes in the parameters of the complex impedance due to destructive processes caused by dehydration and putrefaction processes.

*Data processing and analysis*. On the basis of experimental data obtained at three frequencies modeled frequency and phase response and their changes during losing of freshness and appearance of destructive processes.

*Discussion and conclusions*. In fresh and stale fruit modulus and phase of the impedance at low and high frequencies have characteristic differences. But this is especially evident on the phase-frequency characteristic, which can be seen that the value of the phase with the loss of freshness at low frequency decreases and increases at high more than twice during one week. Therefore, to assess the functional state of fresh and stale products we suggest use phase portraits of phase response.